\documentclass[11pt]{report}
\usepackage[centertags]{amsmath}
\usepackage{amssymb,SIunits}
\newcommand{\un}[1]{\mbox{\unboldmath$#1$}}
\renewcommand{\b}{\null}
\newcommand{\ub}{\null}
\begin{document}
\begin{titlepage}
\begin{center}
\textbf{\LARGE Source Term in Einstein's Field Equations}

\bigskip\bigskip\bigskip\bigskip\bigskip
\begin{minipage}[t]{.37\linewidth}
G. Sudhakaran \\
Chethana ,
H \& C Road\\ 
West Mundakkal\\ 
Kollam - 691 001 \\
Kerala, India \\
Email: shibis@usa.net
\end{minipage}
\hspace*{1cm}
\begin{minipage}[t]{.45\linewidth}
C. Sivaram\\ 
Indian Institute of Astrophysics\\
Bangalore, India. \\
Pin - 560 034.
\end{minipage}
\end{center}

% \vspace*{1cm}
% This is the detailed version of the paper presented by the first author at 
% International Conference on Gravitation and Cosmology organised by IUCAA, 
% Pune (Dec. 13-19, 1995). A different version (CONCEPT OF NON 
% GRAVITATIONAL NATURE OF MATTER) of the same paper presented and 
% published in the proceedings of conferences \cite{one,two,three}. 

% \vspace{1cm}

% The abstract of this paper `Cosmic Repulsion in Presence of Matter' is
% published in GR15 [conference organised by IUCAA, Pune] ``Abstracts of
% Plenary Lectures and Contributed Papers'' (Dec. 16-21, 1997), see pp : 
% 136-137. 
\end{titlepage}

\section*{Keywords}
 Repulsive response of spacetime, Effective pressure, Upper limit of 
temperature. Repulsive fluid, Generality of singularity theorems.

\section*{Abstract}
A basic problem that confronts the standard cosmological
models is the problem of Initial Singularity characterised by infinite
material density, infinite temperature and infinite spacetime
curvature. The inevitable existence of such a phase of the universe
may be considered to be one of the major drawbacks of Einstein's field
equations. To some extent inflation models ameliorate this. In the
present work we postulate that whenever matter (radiation) arises in
flat spacetime, it introduces curvature and causes a repulsive
interaction to develop. A modified energy momentum tensor is
introduced towards this end, which invokes the temperature and
entropy. This redefinition of the energy momentum distribution when
applied to the early universe dynamics has the effect of producing a
non-singular initial behaviour. The repulsive interaction introduces
some features of an accelerating universe consistent with present
observations of high redshift supernovae, etc.
\section*{1. Introduction} 

 In this paper we postulate that whenever matter is introduced in flat 
spacetime, it is stimulated to respond in the following manner;
\begin{enumerate} 
\item spacetime curvature corresponding to $Gij$ (Einstein tensor) is produced and 
\item spacetime responds to temperature and entropy of material
  distribution in such  a way that it develops a peculiar kind of repulsive interaction. 
\end{enumerate}

Thus gravitational field equations in presence of matter could be
written in the  form 
\[-K Tij   =  Gij + G^*ij\]
$Gij$ is Einstein Tensor and $G^*ij$ is the tensor corresponding to
repulsive interaction  of spacetime. 

The very presence of temperature and entropy of material distribution
stimulates spacetime to develop the repulsive interaction. Thus we
assume the existence a repulsive fluid having emt  $Sij$ such that

\begin{align*}
G^* ij    & =  -KSij \\
\intertext{where $Sij$ is assumed to be of the form \cite{one,two,three,newfour}}
Sij &= \dfrac{f(t)\phi\theta U_i U_j}{C^2}  - \alpha(t) g_{ij} \mbox{ with the restriction $S^{ij}_{; j} = 0 $} 
\end{align*}

$f(t)$ is called the degree of repulsive interaction to the presence
of entropy and temperature of material distribution. It is
dimensionless variable. $\phi(t)$ is the entropy density and
$\theta(t)$ is  temperature of the material distribution. $U_i$ is
the 4-velocity vector. $C$ is the velocity of light in freespace.
$\alpha(t)$ is a cosmic scalar field depending on cosmic time t.
When $\theta$ = Zero Kelvin\ we assume that $\alpha(t) = 0 $.

Hence the modified form of gravitational field equations take the form 

\begin{align*}
 Gij  &=  -K (Tij - Sij)  = -K Tij \text{ (effective)}
\text{i.e.,} 
         Gij  &=  -K \bigl[Tij + (-Sij)\bigr] 
\end{align*}

$Tij$ is the total energy tensor. 
$Sij$ is the energy tensor corresponding to the repulsive interaction. 
     $Tij$ (effective) represents the effective energy tensor. 
\begin{align*}
  \bigl[T_{ij} \text{ (effective) } - \dfrac{1}{2}\ g_{ij} T \text{(effective)}\bigr]
  U^i U^j &\ge 0 \\
\text{implies } 
\bigl[T_{ij} - \dfrac{1}{2}\ g_{ij} T\bigr] U^i U^j \ge \alpha(t) +
 \dfrac{1}{2}\dfrac{f(t)\phi\theta}{C^2} &\ge 0
\end{align*}

Details regarding the above inequality are being carried out by us. 

%\section*{Introduction} 
\section*{2. Basic Assumptions}
\begin{enumerate} 
\label{asone}\item[(A1)] For the sake of convenience, we restrict to
study the dynamics of the universe for the case when cosmic time is
non-negative. But however without loss of generality one can extend to
study the nature of the universe for cosmic time $t$
less than zero. It could be analysed if one assumes the discontinuity of
  Hubble parameter at $t=0$. The universe is already in the state of
  BEING at $t=0$ for there is NO geodesic singularity at $t=0$. 
\label{astwo}\item[(A2)] To begin with a peculiar kind of blackbody radiation field
  exists and only this radiation dominates throughout the radiation
  dominated era for $t \ge 0$, the entropy density is given by
  ($\frac{4}{3})a \theta^3$ throughout the radiation
  dominated era. Here \b{a} is the radiation constant. In this model
  only thermal entropy is considered. During evolution the total
  entropy of the universe is conserved. The law $\theta R $ is constant
  is true for all $t \ge 0$, $\theta$ is the temperature and R is the
  cosmic scale factor.

The expression \b $ C^{'} = \dfrac{\un{\phi\theta} R^4}{C^2}$ \unboldmath = constant is true for t $\ge$ 0. 
%\end{enumerate}
\label{asthree}\item[(A3)] There exists \b{R}(0) $\not=$ 0 at $t = 0$ and \b{R}(0) is
  finite and greater than zero.  $\theta(0)$ is the finite initial
  temperature at cosmic time t = 0. Thus the law $\theta$R = constant,
  gives $\theta$\b{Y} = $\theta$(0) and \b $ Y =
  \dfrac{R}{R\un{(0)}}$\ub.
\label{asfour}\item[(A4)] The form of energy tensor\cite{four} Tij is assumed to be of the form 
          
       \b \[ Tij       = \bigl[\un{\rho} + \dfrac{p}{C^2}\bigr]Ui Uj -
       \dfrac{p}{C^2}\ub g_{ij}\] for all time t $\ge$ 0

In matter dominated era, this form of Tij still holds with p becoming
       vanishingly small. 

Robertson Walker metric\cite{five} is used.  Signature of a spacetime is  -2.
For comoving co-ordinate system Ui = (1, 0, 0, 0). The effective tensor              Tij (effective) causes Gij 
        \b \[  Gij = -K Tij \text{(effective)} =  -K (Tij - Sij)\] \ub 

\label{asfive}\item[(A5)]  There exist relations $\rho = 3 \dfrac{p}{C^2}$ and $\rho_e =
3\dfrac{p_e}{C^2}$ which are true for all $t\ge0$ in radiation
dominated era, $\rho$ is total material density. 
 p = total pressure. 
$\rho_e$ is the effective density causing gravitation and $p_e$ is the effective pressure of material distribution. 

          \b \[ \rho_e = \rho - \dfrac{f(t)\phi\theta}{C^2} \ge 0,
          \dfrac{p_e}{C^2} = \dfrac{p}{C^2} - \alpha(t) \ge 0 \] \ub

In matter dominated era, $p \approx 0, p_e \approx 0$ and $\alpha \approx 0$

\label{assix}\item[(A6)] The Hubble parameter at t = 0 is given by 
  \b \[ H(0) = \bigl[(\dfrac{8}{9}) \pi G \rho(0)\bigr]^{1/2} \] \ub

Here $\rho(0)$ is the total density at t = 0 and it is given by \b
$\rho(0) =   \dfrac{a\theta(0)^4}{C^2}$ \ub

G is Newtonian gravitational constant `\b{a}' is the radiation constant. 
\label{asseven}\item[(A7)] There exists a temperature $\theta_T$ at which transition from
radiation domination to matter domination takes place. At this
temperature, $\rho, \rho_e$, Y, \.{Y} and f(t) are assumed to continuous. Dot shows differentiation with respect to time. 
\label{aseight}\item[(A8)] This model becomes identical to existing Cosmological model of the early 
universe when Y = 1.6824079...  It is the least value of Y at which our model 
becomes identical to the traditional cosmological model of early universe. We 
assume that it occurs at cosmic time\cite{six}  $t = 5.39 \times
10^{-44}$ second, the Planck time. 
\label{asnine}\item[(A9)] To analyse the behaviour for $t<0$, we assume
 discontinuity of the Hubble parameter $H(t)$, at $t=0$. However
$H(0)t$ is always Non negative quantity in the solution given by equation
(\ref{sixteen}) of this MS. So for negative $t$, $H(0)$ has to be negative. So we
can give the interpretation of the contracting phase of the universe
for $t<0$ (indicating a cosmological blue shift). Thus there is no
geodesic singularity at $t=0$. This can then be interpreted (i.e. the
discontinuity in $H(0)$) as a `bouncing' (or reflection) of the
universe at $t=0$.   $t > 0$, then corresponds to the expanding
phase. Here we put 
\begin{align*}
\lim_{t \to \overline{O}} H(t) &= -H(0) \text{ and }\\
\lim_{t \to   \overset{+}{O}} H(t) &= +H(0) 
\end{align*}
\end{enumerate}
\subsection*{3. Repulsive Interaction of Spacetime and Sij term in the field equations}

Einstein field equation \cite{seven} in its classical form 

         \[ Gij = - K Tij , K = \dfrac{8\pi G}{C^2} \]
         
         connects the dialectical categories content and form. Gij,
         the Einstein tensor represents the form explicitly manifested
         as the curvature of spacetime and Tij represents the material
         content causing the manifestation of the form. The content
         and form are connected by the coupling constant K, the
         relativistic gravitational constant. 

         Repulsive gravity can occur in unusual situation and in
         recent years has been invoked in connection with inflation.
         In general relativity both pressure and energy density
         contribute to the gravitational field and inflation invokes a
         negative pressure as implied by a positive cosmological
         constant (induced by a change in vacuum energy) to drive the
         initial expansion of the Universe (as opposed to the usual
         tendency to collapse).  Gravitational binding energy is
         negative and thus reduces the effective energy of the system
         and can make it vanish under certain circumstances. We invoke
         thermal effects in gravitational field which can reduce the
         system energy. In this context, we note the work of
         references \cite{eight}, \cite{nine} where it is shown
         that thermal finite temperature corrections to the energy
         momentum tensor are of the form
\[ \theta^{\mu v} = T^{\mu v} -
         (\dfrac{2}{3})\alpha\pi(\dfrac{T^2}{E^2})\delta_0^{\mu}
         \delta_0^{v} T^{00} \]

At finite temperature, the Minkowski vacuum is replaced by a thermal
bath and note that it has the effect of `diluting' $T_{\mu v}$ and
indeed the expression for acceleration of a particle (mass m) has a
term of $\sim \dfrac{2}{3} (\dfrac{T^2}{m^2})$ of opposite sign to
gravitational acceleration, suggesting thermal induced repulsion. This
is the result of the detailed finite temperature calculation \cite{ten}.

          Here we assume that the response or reaction of spacetime
          consists of two parts
\begin{enumerate} 
\item Producing curvature of spacetime or Einstein tensor Gij 
\item Producing a repulsive response of spacetime (repulsive
interaction) induced by  the presence of entropy and temperature of
material distribution. 
\end{enumerate}

Now gravitational field equation assumes the form 

       \b \[Gij + G^* ij = - K Tij, (K  = \dfrac{8\pi G}{C^2}) \] \ub

Einstein identified \b $G^*_{ij}  = \wedge\ g_{ij}$ \ub 
Here we assume $G^*ij$ to be of the form 
\begin{align*}
G^*ij &= -K Sij \text{ with}\\ 
Sij &=  \dfrac{f(t)\phi\theta}{C^2}Ui Uj - \un{\alpha}(t) g_{ij}
\end{align*}       

If we put f(t) = 0 and $\alpha(t)$ = a constant, we get back Einstein's field equations with $\wedge$ term. 

We do have Einstein's field equations with $\wedge$ - term
                   \[ -K T_{ij} = G_{ij} + \wedge g_{ij}\]
          i.e., matter $\longleftrightarrow$ (Attractive Interaction) + (Repulsive Interaction) 

          It shows that the concept of repulsive interaction was already there in the 
mind of Einstein. In this paper even though $S_{ij}$ is a particular form, it is a bit more 
general than just $\wedge g_{ij}$ term. And the result it generates even in perfect fluid 
approximation does not violate any known physical laws and hence may have some 
physical relevance in cosmology. Thus the existence of $Sij$ is physically plausible. 
We have the restriction $S^{ij}_{; j} = 0 $ and effective pressure,
effective density and $\alpha(t)$ we introduce are Non-Negative quantities. It does not lead to violation of 
energy condition. 

          Except horizon problem and dark matter problem, some fundamental basic 
issues are addressed well in a very natural way without imposing any other extra 
additional suppositions other than the introduction of the repulsive term $Sij$, the 
simplest form of which was already introduced by Einstein. 

          In this paper we hypothesize that introduction of matter is flat spacetime (in 
the early universe), apart from causing a curvature (given by Einstein tensor) would 
also give rise to another effect, i.e., spacetime acquires a black body like 
temperature \textit{due to the curvature} and becomes filled with radiation with a 
corresponding entropy density and a negative pressure leading to repulsive action 
making the spacetime to expand. We choose a particular form for the stress tensor 
of this radiation field. (It could be thought to be a time depended generalisation of 
modified varying cosmical constant term). Such an additional term in the field 
equations can be justified in several physical context, e.g., in the framework of 
quantum gravity and super gravity. The modified field equations are found to have 
solutions which can ameliorate several problems in conventional cosmology such 
as Initial Singularity, the Flatness problem etc. The model also smoothly joins onto 
the radiation era and to present matter dominated phase of the universe and have 
some definite predictions at the present epoch to differentiate it from conventional 
cosmology. 
\section*{4. Justification for the repulsive interaction}
 
          Introduction of matter of density $\rho$ in flat spacetime causes
          it to curve.  This  is a response of spacetime in the sense
          of elastic medium following Satcharov; the  
gravitational constant being constant of elasticity. 

Curvature scalar $\underset{\sim}{R} = K \rho $(we use $\underset{\sim}{R}$ to
distinguish from scale factor $R$), $K = \dfrac{8\pi G}{C^2}, \dfrac{1}{K}$ = Elastic constant (very high), so it needs a high $\rho$ to effectively curve spacetime  as effective elastic constant $\dfrac{1}{K}$ is very high. 

Any curved spacetime with average curvature $\underset{\sim}{R} $, would
by Quantum Gravity  effects, acquire a black body temperature given by 

\newcommand{\Runder}{\underset{\sim}{R}}
\begin{alignat*}{2}
\theta &= \dfrac{\hslash c}{k_B}\sqrt{\Runder}, \hspace{1in} &\sqrt{\Runder}
          &\sim 10^{33} \centi\reciprocal\meter \\ 
          \text{ If } \theta &\sim 10^{32} K, &\Runder &\sim
          \dfrac{C^3}{\hslash G} = 10^{66} \centi\rpsquare\meter 
\end{alignat*}

$R$ changes as spacetime expands. So $\theta$ changes
$\dfrac{\theta}{\sqrt{\Runder}} $ = constant i.e., $\theta R$ is a 
constant for adiabatic expansion. So $S_{ij}$ for radiation field
could be expressed as

       \[  S_{ij} = f(t) \phi\dfrac{\theta}{C^2} U_i U_j - \alpha(t) g_{ij}\]\\ 
          % $f(t)$ is called degree of repulsive interaction \\
%           $\phi(t)$ is the entropy density of material distribution \\
%           $\theta(t)$ is the temperature of material distribution \\
%           $\alpha(t)$ is a cosmic scalar field \\ 
%           $U_i$ is the 4-velocity vector 

This radiation field could consist of ultra relativistic vector particles (of spin 1) the 
so called gravi photons of supergravity theories. It is postulated to be a mediating 
boson particle in addition to graviton and is expected to mediate a repulsive force 
field. Being massive vector field its velocity could be less than $C$, so it behaves like 
a fluid with the above form of $S_{ij}$.  Lorentz invariance would constrain $S_{ij}$ to have 
its natural form as above. 

Thus we have the modified form of Einstein field equations. 
      
\begin{equation} 
\begin{split}
\label{one}Gij       &= -K (Tij - Sij) \\
          &= -K Tij \text{ (effective)}
\end{split}
\end{equation} 

This would reduce to Einstein's field equations when $\theta = 0$ Kelvin.
We assume that $\alpha  = 0$ when $\theta$ = zero Kelvin. 
-Sij corresponds to repulsive interaction. Its existence does not violate the 
principle of equivalence. \textit{We imposed the restriction} 
\begin{align*}
       S^{ij}_{; j}   &=  0
\intertext{Already we have }
          G^{ij}_{; j}   &=  0 
\intertext{and} 
         T^{ij}_{; j}    &=  0
\end{align*}
It is simple approach to remove the Initial Singularity.  Now if we calculate the quantity

\[ Ui [T^{ij} - S^{ij}]_{; j} = 0 \] 
for an elemental volume being proportional to $R^3$ in
Robertson Walker spacetime, we get

\begin{align}
\label{two}d (\un{\rho_e} R^3) + p_e (d\dfrac{R^3}{C^2})
       &= -d (\un{\alpha} R^3) 
\intertext{where $\rho_e$ is the effective density} 
\label{three}\rho_e = \rho -  \bigl[f(t) \phi\dfrac{\theta}{C^2}\bigr] &\ge  0
\\        \label{four}\dfrac{p_e}{C^2} = \dfrac{p}{C^2} - \alpha(t) &\ge 0        
\end{align}

\noindent $p_e$ is the effective pressure of material distribution.\\
$\rho_e = 0 \text{ at } t = 0 $ implies
\begin{equation}
\begin{split}
\rho(0) &=  \dfrac{f(0)\phi(0)\theta(0)}{C^2} \\  
\rho(0) &=  \dfrac{a\theta(0)^4}{C^2} \\
\label{five} \text{ and }\phi(0) &= \dfrac{4a\theta(0)^3}{3}   \\
\text{ implies } f(0)  &=  \dfrac{3}{4}
\end{split}
\end{equation} 
$p_e(0) = 0$  at $t = 0$  implies 
\[ \alpha(0) = \dfrac{p(0)}{C^2} = \dfrac{\rho(0)}{3}\]

$U_i T^{ij}_{; j}  = 0 $\text{ yields }
\begin{equation}
\label{six}
 d (\rho R^3) + \dfrac{p(dR^3)}{C^2} = 0
\end{equation}
since $\rho  = \dfrac{3p}{C^2}$ eqn (\ref{six}) gives \\
\begin{equation}
\label{seven}\rho R^4  =  \rho(0) R^4(0) = C^* = \text{ constant} 
\end{equation}
$U_i  S^{ij}_{; j} = 0 $ gives 
\begin{equation}
\begin{split}
  \label{eight}\alpha(t)  &= C^{'}\bigl[f(t) R^{-4} + 3 \int f(t) \dfrac{R^{-4}dR}{R}\bigr] + \text{constant}     \\
  \text{Where } C^{'} &= \dfrac{\phi\theta R^4}{C^2} = \text{ constant
  for } t \ge 0.
\end{split}
\end{equation}
Now  eqn (\ref{three}) and eqn (\ref{four}) along with the expressions 
$\rho = \dfrac{3p}{C^2}$  
 and $ \rho_e = \dfrac{3 p_e}{C^2} $ give 
\begin{equation}
  \label{nine}\alpha = \dfrac{1}{3} C^{'} (\dfrac{f(t)}{R^4})
\end{equation}
From  eqn (\ref{eight}) and eqn (\ref{nine}) we shall have 
      
\[    f(t) =  \dfrac{f(0)}{\sqrt{Y}},  Y = \dfrac{R}{R(0)}, (1 \le Y
          < \infty) \] 
From  eqn  (\ref{five})
\begin{equation}
\label{ten}  
f(0)  =  \dfrac{3}{4} \text{ hence } f(t) = \dfrac{3}{4\sqrt{Y}}
\end{equation}

 From  eqn (\ref{nine}) and eqn (\ref{ten}) we obtain 

\[ \alpha(t) = \dfrac{1}{3} \rho(0) Y^{-(\dfrac{9}{2})}\]
From eqn (\ref{three}) 
\begin{align*}
          \rho_e R^4 &= \rho R^4 - \dfrac{f(t)\phi\theta R^4}{C^2}\\
                    &=  C^* (1 - \dfrac{f(t) C^{'}}{C^*})\\
          \rho R^4  &=  C^* = \rho(0) R(0)^4 \text{ and } f(0) =
                    \dfrac{C^*}{C^{'}},  f(t) =\dfrac{f(0)}{\sqrt{Y}}
\end{align*} 
Therefore
\begin{align} 
        \rho_e R^4    & =  C^* \bigl[1 - (\dfrac{1}{\sqrt{Y}})\bigr]
        \notag \\
\label{eleven}\text{i.e., }\rho_e Y^4  &=  \rho(0)\bigl[1 - (\dfrac{1}{\sqrt{Y}})\bigr]
%\end{split}
%\end{equation}
\end{align}

From  eqn (\ref{seven})
\begin{align*}
 \rho Y^4  &= \rho(0) \\
\text{Therefore \: }
\rho_e &= \rho\bigl[1 - (\dfrac{1}{\sqrt{Y}})\bigr]\\
\intertext{Y becomes very large implies }
\rho_e &\sim \rho \\ 
\intertext{For a comoving co-ordinate system }
        U^0 &=  1,  U_0 = 1, g_{00} = +1 \\ 
          U^1 &= U^2 = U^3 = 0 \\
\text{Now } 
         T^i_{j} - S^i_{j} &= (\rho_e + \alpha) U^i U_j - p_e C^{-2} (\delta^i_j - U^i U_j)\\ 
\text{we have } 
         T^i_i - S^i_i  &= \alpha \not= 0
\end{align*}
since $ \rho_e = 3 \dfrac{p_e}{C^2} $ in radiation dominated
  era \\ 
Now the field equation (\ref{one}) in Robertson Walker spacetime yield 
\begin{align}
\label{twelve} (\dfrac{8\pi G}{C^2}) (\rho_e + \alpha) &= 3(\dfrac{k}{R^2} +
          \dfrac{\text{\.{R}}^2}{C^2R^2})\\ 
\label{thirteen}(\dfrac{8\pi G}{C^2}) (\dfrac{p_e}{C^2}) &=
-(\dfrac{k}{R^2} + \dfrac{\text{\.{R}}^2}{C^2R^2}) - (\dfrac{2\text{\"{R}}}{C^2 R}) 
\end{align}
Here $k$ is the curvature index. Dot shows differentiation with
  respect to time. Now substituting for  $\rho_e, \alpha $ and  $R$ in 
terms of $Y$ variable and making use of equation
         \[ \rho_e  = 3 \dfrac{p_e}{C^2}\]
Eqn (\ref{twelve}) { and 
          eqn} (\ref{thirteen}) { yield after integration}\\
\begin{equation}
 \label{fourteen}\biggl[\dfrac{dY}{dt}\biggr]^2 = (8\pi\dfrac{G}{3}) \rho(0)(Y^{-2} -
          (\dfrac{2}{3}) Y^{-(5/2)}) 
\end{equation}
The constant of integration vanishes since we assume that 
\[   H(0) = (\dfrac{8\pi G \rho(0)}{9})^{1/2} \] 
is the  Hubble parameter at  $t = 0$.
Thus Hubble parameter in radiation dominated era 
\begin{equation}
\label{fifteen}H(t) = \pm H(0) (3 Y^{-4} - 2Y^{-9/2} )^{1/2} 
\end{equation}

$H(t)$ is taken to be negative when $t<0$ showing blue shift. $H(t)$
is taken to be positive  when $t > 0$ showing red shift. 

From  eqn (\ref{eleven}) it is clear that  $\rho_e$  increases
  for $ 1 \le Y  < \dfrac{81}{64}$  and it decreases for 
  $\dfrac{81}{64} < Y < \infty, $  becomes maximum at $ Y = \dfrac{81}{64}$\\
On integrating eqn (\ref{fourteen}) we shall have 
\begin{multline}\label{sixteen}
\Delta \biggl[(\dfrac{1}{2}) Y^{7/4}  + (\dfrac{7}{18}) Y^{5/4} \\
+ (\dfrac{35}{108})(Y^{3/4} + Y^{1/4})\biggr] + (\dfrac{35}{162}) ln
|Y^{1/4} + \Delta|  =  \big[\sqrt{3}\big] H(0) t + \mu 
\end{multline}
$\mu$ is the constant of integration and  $ \Delta = ( \sqrt{Y} - \dfrac{2}{3})^{1/2}$ \\
At $ t = 0, Y = 1 $ so that $ \mu = 0.985872474...$\\
when $Y$ becomes appreciable $ \dfrac{1}{2} Y^2 $ dominates 
  LHS of eqn (\ref{sixteen}) and we shall have \\
\begin{equation}
\label{seventeen} Y = (2\sqrt{3} H(0))^{1/2}\sqrt{t} 
\end{equation}
Equation (\ref{sixteen}) collapses to the above equation when $ Y =
1.6924079148.....$ 
According to our assumption this occurs when $ t = 5.3903687 \times
10^{-44} $\second.  
Hence we obtain 
\[ \theta(0) = 1.107 \times 10^{32} \kelvin\] 
\begin{align}
\intertext{From equation  (\ref{seventeen}),}
\label{eighteen} R(t) &= R(0)(2\sqrt{3} H(0)t)^{1/2}\\
\intertext{From equation (\ref{eleven})  we get} 
 \label{nineteen} \rho_e &=  \dfrac{(\rho(0)(H(0)t)^{-2}}{12} \\
\label{twenty}  &=  \dfrac{3}{32\pi G}\:\dfrac{1}{t^2} \\
\intertext{From $\theta Y = \theta(0)$, we get}
 \label{twentyone} \theta &\approx \theta(0) (2\sqrt{3} H(0))^{-1/2}
                    \dfrac{1}{\sqrt{t}} \\  
\intertext{Substituting the value of $H(0)$ we get}  
\label{twentytwo}\theta &\approx \dfrac{{\biggl[\dfrac{3C^2}{32\pi G
          a}\biggr]}^{1/4}}{\sqrt{t}} \\ 
\label{twentythree}\text{i.e., }    \theta &\approx \dfrac{1.52 \times 10^{10}}{\sqrt{t}}  \kelvin
\end{align}
Expressions (\ref{eighteen}), (\ref{nineteen}) and (\ref{twentyone})
are similar to those expressions in the existing  cosmological theory of early universe.

The deceleration parameter in radiation dominated era 
\begin{align*}
          q(t)      &=  \dfrac{\sqrt{Y} - \dfrac{5}{6}}{\sqrt{Y} - \dfrac{2}{3}}\\
          q(0)     & =  \dfrac{1}{2}\\
\intertext{and $q(t)$ tends to 1 as $Y$ tends to $\infty$.}
\intertext{Density parameter in the radiation dominated era is found to be }
          \Omega (t)     &=  \dfrac{\rho_e}{\rho^*},  \rho^*  =
          \dfrac{3H^2}{8\pi G} \\
\intertext{$H$ is the Hubble parameter in the radiation dominated era,  we get  }
   \Omega(t)      &=  \dfrac{1-(\dfrac{1}{\sqrt{Y}})}{1- (\dfrac{2}{3\sqrt{Y}})}
\end{align*}

The flatness problem in the traditional standard Cosmological
  picture may be reviewed in the light of this model. In this model
  the curvature index k vanishes identically in radiation dominated
  era (See Appendix III).

\section*{5. Matter dominated era}
For matter dominated era we have $p \approx 0, p_e \approx 0$
yielding $\alpha \approx 0$ \\
Thus equations (\ref{twelve}) and (\ref{thirteen}) read 
\begin{align}
\label{twentyfour}
\dfrac{8\pi G}{C^2} \rho_e  &=  3 (\dfrac{k}{R^2} +
\dfrac{\text{\.{R}}^2}{C^2R^2})\\ 
\label{twentyfive}\text{and  } 0 &= -3\biggl[\dfrac{k}{R^2} +
\dfrac{\text{\.{R}}^2}{C^2R^2}\biggr] - \biggl[6\dfrac{\text{\"{R}}}{C^2R}\biggr] \\ 
\intertext{On adding (\ref{twentyfour}) and (\ref{twentyfive})}
\label{twentysix}
\dfrac{8\pi G}{C^2} \rho_e  &=
-(\dfrac{6\text{\"{R}}}{C^2R}) = -(\dfrac{6\text{\"{Y}}}{C^2Y})\\  
\label{twentyseven}\text{But } \rho_e Y^3 &= \rho Y^3 -
\dfrac{f(t)\phi\theta Y^3}{C^2} 
\end{align}
From (\ref{two}) and (\ref{six}) we shall have
\begin{align*}
          d(\rho_e Y^3) &= 0 \text{ and }  d(\rho Y^3) = 0 \text{ for }
          p_e = 0, p = 0 \\
\text{ and } \alpha &= 0 \text{ in the matter dominated era.} \\
\intertext{Therefore }  
\rho_e Y^3 &= B_1 = \text{ constant and } \rho Y^3 =
B_2 = \text{ constant} \\
\intertext{Thus eqn (\ref{twentyseven}) suggests } 
(\dfrac{f(t) \phi\theta}{C^2})
(\dfrac{Y^4}{Y}) &= B_3 = \text{ constant}\\
          f(t) &= \dfrac{B_3Y}{A_3}\\
          \text{where } A_3 &=  \dfrac{\phi\theta Y^4}{C^2}  = \dfrac{4}{3} (\dfrac{a \theta^4(0)}{C^2})
\end{align*}
Then eqn (\ref{twentyseven}) takes the form $ B_1 = B_2 - B_3$.
Continuity of $ \rho_e, \rho, Y,$ {\.{Y}}  and $ f(t)$ at  $\theta_T $
help us to determine the constants $ B_1, B_2$  and $B_3$. Thus 
%\begin{equation}
%\begin{split}
\begin{align} 
         B_3  &=  {\biggl[\dfrac{\theta_T}{\theta(0)}\biggr]}^{3/2} \rho(0), B_2  =
          \dfrac{\theta_T}{\theta(0)} \rho(0) \notag\\ 
          B_1  &=  \rho(0) \dfrac{\theta_T}{\theta(0)}\biggl[1 -
          \sqrt{\dfrac{\theta_T}{\theta(0)}}\biggr] \notag\\
\text{we have  } 
\label{twentyeight}\rho_e &= B_1 Y^{-3} \\
\rho &= B_2 Y^{-3} \text{ and }\notag\\
  f(t) &= \dfrac{3}{4}{\biggl[\dfrac{\theta_T}{\theta(0)}\biggr]}^{3/2} Y \notag
%\end{split}
%\end{equation}
\end{align}

since $\theta_T \ll \theta(0), B_1 \sim B_2 $ showing that $\rho_e
\sim \rho$ \\ 
Making use of equation (\ref{twentysix}) and
equation (\ref{twentyeight}) one obtains after integration
\begin{align}
H^2(t)  &= H^2(0) \biggl[\dfrac{\theta_T}{\theta(0)}\biggr]
\biggl\{3\biggl[1 - \sqrt{\dfrac{\theta_T}{\theta(0)}}\biggr] Y^{-3} +
{\biggl[\dfrac{\theta_T}{\theta(0)}\biggr]}^{3/2} Y^{-2}\biggr\}\notag
\\
%H(t) &< 0 \text{ for } t < 0 , H(t) > 0 \text{ for } t \ge 0
\intertext{Thus}
\label{twentynine}{H(t)\choose (matter)} &= \pm \sqrt{\text{RHS of the above  equation}}
\end{align}

$H(t) < 0 \text{ for } t < 0 ,\quad H(t) > 0 \text{ for } t > 0 $

Here constant of integration is obtained on the assumption of
  continuity of $H(t)$ at $\theta_T$. {$H(t)\choose \text{(matter)}$}  is the Hubble parameter in matter dominated era.

Now the present value of  $Y = Y_N = \dfrac{\theta(0)}{\theta_N}
= \dfrac{R_N}{R(0)}$

Substituting this value of $ Y_N$  in expression (\ref{twentynine}) and
  evaluating  $\theta(0)$ \\
\begin{align*}
\text{We  obtain  } \theta(0)   &= \dfrac{\theta_T\biggl[\dfrac{\theta_T}{\theta_N}-3\biggr]^2}{{\biggl[\dfrac{\theta^*}{\theta_T}-3\biggr]}^2}\\
\text{Where } \theta^*  &=  \dfrac{9H^2_{N}C^2}{\theta^3_{N}8\pi G a}
\end{align*}

$H_N$ = present value of Hubble parameter, $\theta_T = 4000$
  \kelvin, the transition temperature  from radiation domination to
  matter domination and \hbox{$\theta_N = 2.7$ \kelvin.} 

Since the exact value of $H_N$ is unknown the determination of
  $\theta(0)$  using the above expression is inappropriate.

Now consider the assumption (A8), it is well accepted that the classical limit upto which one can march
towards $t = 0$ is the Planck time
$t\approx5.34\times10^{-44}\sec$. All the cosmical parameters shall
have classical meaning only from $t\approx5.39\times10^{-44}
\sec$. And the variable $Y$ becomes the classical expression
\begin{equation*}
Y = {\biggl(2\sqrt{3} H_{(0)}\biggr)}^{\dfrac{1}{2}}\sqrt{t}
\end{equation*}

When $Y$ lies in the range

\begin{equation*}
1.6924079\cdots \le Y < \text { very large number }
\end{equation*}

Hence the least value of $ Y = 1.6924079\cdots$. It occurs when
$t\approx10^{-44}\sec$ (See Appendix II), and make use of the equation 
\begin{align*}
          Y  &=  (2\sqrt{3} H(0))^{1/2} \sqrt{t}\\ 
\text{i.e., }     Y &=  \biggl[2\sqrt{3} (\dfrac{8}{9} \pi Ga C^{-2})^{1/2}\biggr]^{1/2} \theta(0) \sqrt{t} \\
\text{When } Y &= 1.69240...,  t = 5.39036......... \times 10^{-44}\second. \\
\text{So that } \theta(0)  &= 1.10677... \times 10^{32} \kelvin
\end{align*}
Deceleration parameter in matter dominated era takes the form
\begin{align}
\label{thirty} 
q(t) &= \dfrac{1}{2}\biggl[1 + (\dfrac{1}{3}) {\eta_T}^{3/2}  Y (1 -
\sqrt{\eta}_T)^{-1}\biggr]^{-1} \\
\text{where } 
{\eta}_T &= \dfrac{\theta_T}{\theta(0)}, q < \dfrac{1}{2} \notag
\end{align}
Substituting for $\rho_e$  and  $H^2$ from eqn (\ref{twentyeight}) and  eqn (\ref{twentynine}) in eqn (\ref{twentyfour}) we obtain 
\[k = -1 \text{ and } R(0) =
\dfrac{C}{H(0)}(\dfrac{\theta(0)}{\theta_T})^{5/4}\] 
R(0) being the scale factor at   t = 0 (See Appendix IV).

The density parameter in the matter dominated era assumes the   form 
\[ {\Omega(t) \choose (\text{matter})} = \biggl[1 + (\dfrac{1}{3})
         {\eta_T}^{3/2}  Y (1 - \sqrt{\eta}_T)^{-1} \biggr]^{-1} \]
The present value of density parameter is obtained by putting
\[ Y = Y_N = \dfrac{\theta(0)}{\theta_N}\]
We have 
\[\Omega_N = \biggl[1 + (\dfrac{1}{3}) {\eta_T}^{3/2}  Y_N (1 -
          \sqrt{\eta}_T)^{-1} \biggr]^{-1} \]
which is less than 1, but very nearly equal to 1.
\[ \Omega_N =  \dfrac{\rho_{eN}}{\rho^*_{N}} \]
$\rho_{eN}$ = present value of effective density. $\rho^*_N$ =
 present value of critical density.

The expression for $\Omega_N$ implies that $\rho_{eN}$ is less
  than ${\rho^*_{N}}$ and very nearly equal to ${\rho^*_{N}}$. That is the
  effective density causing gravitation is very near to the present
  value of critical density but less than that.
\text{We have }  
\begin{align*}
  (1 - \Omega_N)  &=   0.211 \times 10^{-14}\\
  \theta(0)     &=  1.10677 ... \times 10^{32} \kelvin\\
  H(0) &= 1.533913... \times 10^{42}
  \reciprocal\second, R(0) = 6.974818 \metre \\
  H_N &= 0.61 \times 10^{-18} \reciprocal\second, \rho^*_N = 0.6657
  \times 10^{-27} \kilo\gram \meter\rpcubed
\end{align*}
\section*{6. Conclusion} 
Initial singularity is effectively removed at the instant t = 0, the
beginning of this expanding phase of the universe. The Standard
Friedmann model is described as the big bang model for the universe.
It implies that the universe began in an initial singularity. If the
universe is a closed one it will end in future singularity. Standard
Cosmological models lead to the conclusion that the universe began
from a singular state - a state with an infinite high density of
matter. In our model Initial Singularity is avoided by the presence of
term $Sij$ in the field equations. There exists an upper limit of
temperature $\theta(0) \approx 10^{32} $ Kelvin at which the effective
density $\rho_e$ vanishes.  And when $Y = 1.6924... $ at $t = 5.390
... \times 10^{-44} $ sec; this theory coincides with existing
Cosmological theory of early universe. From the expression for density
parameter in the Radiation dominated era it is clear that there is
\textit{no flatness problem} in this model.  Curvature index vanishes
identically in radiation dominated era.

          Density parameter in matter dominated era shows that effective density is 
nearly equal to critical density but bit less than critical density. Curvature index in 
matter dominated era is found to be -1 showing that the
\textit{3-space is hyperbolic}. Thus  
knowing  $\theta_N = 2.7$ \kelvin,  $\theta_T = 4000 $ \kelvin\ and
$\theta(0) \approx  10^{32} $ \kelvin\ a lot of 
information flows out of the theory. This is a simplified model of the universe. 

In conclusion we have a model leading to $\Omega_N$ very close to 1,
and an accelerating phase (due to k = -1). The latter is consistent
with current observational results based on impressive data involving
luminosity evolutions of high red shift supernovae which strongly
suggest an accelerating phase (q = -1). Refs \cite{eleven},
\cite{twelve}.

The fact that $\Omega_N$ is very close to one, is supported by very
recent observations of the \textbf{BOOMERANG} project of the cosmic
microwave background as reported in the April 27, 2000 of nature \cite{thirteen}.

Eventhough the term $S_{ij}$ is purely arbitrary and adhoc it
challenges the generality of singularity theorems since $T_{ij}
\mbox{ (effective) } = (T_{ij}-S_{ij})$ fits well within the
conceptual structure of GTR without violating any known physical laws.  
\newpage
\centerline{\textbf{\Large Appendix I}}

\bigskip

This is to show how we could get the number Y = 1.6924079....

From Equation (16) of the manuscript we have 
\begin{multline*}
\Delta \times \dfrac{1}{2} Y^{7/4} + \biggl[\Delta\biggl\{\dfrac{7}{18} Y^{5/4} +
\dfrac{35}{108} (Y^{3/4} + Y^{1/4})\biggr\}\\ + \dfrac{38}{162} log_e | Y^{1/4} +
\Delta | - \mu\biggr]  =  \sqrt{3} H(0)t
\end{multline*}
\[\text{i.e., } \Delta \times \dfrac{1}{2} Y^{7/4} + A_1  =
          \sqrt{3} H(0)t \]
 where $A_1$ is the term within the square bracket of the above equation
\begin{align*}
  \mu &= 0.985872474...\\
  \Delta &= (\sqrt{Y} - \dfrac{2}{3})^{1/2} 
\end{align*}
Squaring the above equation and rearranging the terms we shall have 
\[\dfrac{1}{4}Y^4 - \dfrac{1}{6} Y^{7/2} + \bigl[A_1^2 + A_1 \Delta
       Y^{7/4}\bigr]  =  3H(0)^2 t^2 \]
\begin{align*}
\intertext{when}
 \dfrac{1}{4} Y^4 &= 3H(0)^2 t^2 \\
\intertext{then }
\dfrac{1}{6} Y^{7/2} &= A_2 \\
\intertext{where}
A_2 &= A_1^2 + A_1 \Delta Y^{7/4}\\
\text{i.e., }  Y &= (6A_2)^{2/7}
\end{align*}

Now using this equation we can find the range of value of $Y$ for
which this equation is satisfied,  the range being  
        \[  1.6924079...  \le Y < \text{ very very large number}.\]
Hence the least value of Y = 1.6924079... It occurs when $t \sim 10^{-44} \second$.  

\newpage
\centerline{\textbf{\Large Appendix II (Some subtle features)}}

\bigskip
\begin{enumerate}
\item  Our theory does have entirely different structure for $f(t)
  \not= 0 $ in both radiation era and matter dominated era. The word identical may be replaced. One may say 
that there is one-one correspondence between our theory and accepted Big Bang 
theory after $10^{-44}$ sec. The correspondence is such that they appear to be identical 
for $Y \sim t^{1/2}$ in radiation dominated era and $Y \sim t^{2/3}$ in matter dominated era as 
expected by conventional cosmologists as far as the scale factor is concerned. 
\item Initial Singularity is removed in a very natural way. All
  cosmological parameters are mathematically well behaved. There is no
  abnormality in any of the cosmological parameters. Explicit
  expressions are obtained for all of them.  Existence of horizon is
  taken to be an initial condition. This is one of the limitations of
  our theory. \textsc{AND THE TEMPERATURE AT THE BEGINNING OF THIS
    EVER EXPANDING UNIVERSE IS $10^{32}$ \kelvin. THIS COULD BE TAKEN
    TO BE AN UPPER LIMIT OF TEMPERATURE IN KELVIN SCALE}. Some
  theoretical backing up to this upper limit of temperature could be
  found elswhere. Knowing this temperature all other cosmological
  parameters are determined.
\item There is \textit{no flatness problem} in this model 
\item Expected behaviour of scale factor both in radiation dominated era and matter dominated era is contained in this model. 
\item Curvature index vanishes identically in radiation dominated era
  and it becomes   -1 in matter dominated era showing that the 3-space
  is hyperbolic in matter  dominated era. 
\item  If we put $f(t) = 0 $ and $\alpha(t) $ a constant we get back
  Einstein's field equations with $\wedge$ term. 

          Above all it is a very simple model of the universe with just a few 
physically acceptable assumptions. 
\item In effect we have a time-variable $\wedge$ coupled to
  temperature. A time-variable $\wedge$ can also be reconciled with
  most of the observations \cite{fourteen}. A scalar field potential
  of the form $V(\phi)\sim\phi^{-\alpha}, \alpha > 0$, also is
  equivalent to a varying cosmic term \cite{fifteen}. 

% This coupling of
%   $\wedge$ with temperature determines the point when inflation takes
%   over from radiation domination, i.e.,
% \begin{align*}
% \rho_{vac} &= \biggl(\wedge(t)\dfrac{C^4}{8\pi G}\biggr) > \rho_{rad}
% \biggl(=\dfrac{aT^4(t)}{C^2}\biggr)\\
% T(t) &= \hslash C \biggl[\dfrac{\sqrt{\wedge}}{k_B}\biggr] \qquad
% \sqrt{\wedge}\sim \phi^{-1}\\
% (\phi \mbox{ a scalar field})\\
% \mbox{Thus for } \wedge &= \wedge_0f(t),
% \end{align*}
% (so that $\rho$ and P are functions of t as in eqn. for Sij, above)\\
% the solution in the earliest phase is of the form
% \[ R\sim exp\int f(t) dt \]
% (This will solve horizon and flatness problems as usual and have a
% residual repulsive term at the present). See also \cite{sixteen}.
\end{enumerate}

\newpage
\centerline{\textbf{\LARGE Appendix III} }

\bigskip

\subsection*{To show that $k = 0$ in radiation dominated era}

From equation (12) we have

\begin{equation*}
\begin{split}
  3\biggl[\dfrac{k}{R^3} + \dfrac{\overset{\circ}{R}^2}{c^2R^2}\biggr]
  &= \dfrac{8\pi
    G}{C^2}(\rho_e + \alpha)\\
  \dfrac{k}{R^2} + \dfrac{\overset{\circ}{R}^2}{c^2R^2} &= \dfrac{8\pi
    G}{3C^2}(\rho_e + \alpha)\\
  \dfrac{k}{R^2} + \dfrac{\dot{y}^2}{c^2Y^2} &= \dfrac{8\pi
    G}{3C^2}(\rho_e + \alpha)\\
  \text{ Here } \rho_e &=
  \rho_{(0)}Y^{-4}\biggl(1-\dfrac{1}{\sqrt{Y}}\biggr) \text{
    [From eqn (11)] }\\
  \text{ and } \alpha &= \dfrac{1}{3}\rho_{(0)}Y^{-\frac{9}{2}} \text{
    [eqns: (9) \& (10) ]}\\
  \rho_e + \alpha &= \rho_{(0)}Y^{-4}\biggl(1-\dfrac{1}{\sqrt{Y}}\biggr) +
  \dfrac{1}{3}\rho_{(0)}Y^{-\frac{9}{2}}\\
  &= \rho_{(0)}Y^{-4} - \dfrac{2}{3}\rho_{(0)}Y^{-\frac{9}{2}}\\
\dfrac{k}{R^2} +  \dfrac{\dot{y}}{c^2Y^2} &= \dfrac{8\pi G}{3C^2}
  \rho_{(0)}Y^{-4} - \frac{2}{3} \biggl(\dfrac{8\pi G}{3C^2} \biggr)
  \rho_{(0)}Y^{-\frac{9}{2}}\\
\text { But } \dfrac{1}{c^2Y^2}(\dot{Y}^2) &=  \dfrac{8\pi G}{3}
  \dfrac{\rho_{(0)}}{Y^2C^2} \biggl(Y^{-2} - \dfrac{2}{3}
  Y^{-\frac{5}{2}}\biggr)   \text{ [from (14)] }\\
&=\dfrac{8\pi G}{3C^2}\rho_{(0)}  Y^{-4} - \biggl(\dfrac{2}{3}\biggr) \dfrac{8\pi
  G}{3C^2} \rho_{(0)} Y^{-\dfrac{9}{2}}\\
\intertext{ It implies }
\dfrac{k}{R^2} &= 0 
\end{split}
\end{equation*}

i.e. $k$ vanishes identically in radiation dominated era

\newpage
\hspace*{-2cm}
\begin{minipage}{\textwidth}
\centerline{\textbf{\LARGE Appendix IV} }

\bigskip
\subsection*{To show that $k = -1$ in matter dominated era}

From equation (24) we have 
\begin{equation*}
\begin{split}
  \dfrac{8\pi G}{C^2}\rho_e &= 3\biggl[\dfrac{k}{R^2} +
  \dfrac{\overset{\circ}{R}^2}{c^2R^2}\biggr]\\
  \dfrac{8\pi G}{C^2} B_1Y^{-3} &= 3\biggl[\dfrac{k}{R^3} +
  \dfrac{\overset{\circ}{R}^2}{c^2R^2}\biggr] \text{ using (28) }\\
  &= 3\biggl[\dfrac{k}{R^2} + \dfrac{H^2}{C^2}\biggr]\\
  B_1 &= \rho_{(0)} \biggl[
  \dfrac{\theta_T}{\theta_{(0)}}\biggr]\biggl[1-\sqrt{\dfrac{\theta_T}{\theta_{(0)}}}\biggr]\\
\dfrac{8\pi G}{3C^2}\bigg\{\rho_{(0)}\biggl(\dfrac{\theta_{T}}{\theta_{(0)}}\biggr)\biggr[1-\sqrt{\dfrac{\theta_{T}}{\theta_{(0)}}}\biggr]\bigg\}Y^{-3}&=
% %
\dfrac{k}{R^2} + \dfrac{1}{C^2} {H_{(0)}}^2 
\biggl[\dfrac{\theta_{T}}{\theta_{(0)}}\biggr] 
\biggl\{3\biggl[1-\sqrt{\dfrac{\theta_{T}}{\theta_{(0)}}}\biggr]Y^{-3}\\
& \quad +
\biggl[\dfrac{\theta_{T}}{\theta_{(0)}}\biggr]^{\frac{3}{2}}Y^{-2}\biggr\}\\
&=\dfrac{k}{R^2} + \dfrac{8}{9} \dfrac{\pi G}{C^2}\rho_{(0)}
\biggl(\dfrac{\theta_{T}}{\theta_{(0)}}\biggr) 
\biggl\{3 \biggl(1 - \sqrt{\dfrac{\theta_{T}}{\theta_{(0)} } }\biggr)
Y^{-3} \\
&\quad + \biggl(\dfrac{\theta_{T}}{\theta_{(0)}}\biggr)^{\frac{3}{2}}Y^{-2}\biggr\}\\
&=\dfrac{k}{R^2} + \dfrac{8\pi
  G}{3C^2}\rho_0\biggl(\dfrac{\theta_{T}}{\theta_{(0)}}\biggr)\biggl(1-
\sqrt{\dfrac{\theta_{T}}{\theta_{(0)} } }\biggr)Y^{-3} \\
&\quad + \dfrac{8\pi
  G}{9C^2}\rho_0\biggl(\dfrac{\theta_{T}}{\theta_{(0)}}\biggr)\biggl({\dfrac{\theta_{T}}{\theta_{(0)}}\biggr)}^{\frac{3}{2}}Y^{-2}\\
\therefore 0 &= \dfrac{k}{R^2} + \dfrac{8\pi
  G}{9C^2}\rho_0\biggl(\dfrac{\theta_{T}}{\theta_{(0)}}\biggr)\biggl({\dfrac{\theta_{T}}{\theta_{(0)}}\biggr)}^{\frac{3}{2}}Y^{-2}\\
\therefore \dfrac{k}{R^2} &= -\dfrac{8\pi  G}{9C^2} \rho_0
\biggl({\dfrac{\theta_{T}}{\theta_{(0)}}\biggr)}^{\frac{5}{2}}Y^{-2}
\end{split}
\end{equation*}
It implies  $k = -1$  and $R_{(0)}= \dfrac{C}{H_{(0)}}
  \biggl({\dfrac{\theta_{T}}{\theta_{(0)}}\biggr)}^{\frac{5}{4}} $  for $H_0 =
  {\biggl(\dfrac{8}{9} \pi G \rho_0 \biggr)}^{\frac{1}{2}}$
\end{minipage}
% \subsection*{Regarding Assumption A8} 

% Here we mention the number $Y = 1.6824079\cdots$ and the
% least quantum of time $t\approx5.39\times10^{-44} \sec$.

% It is well accepted that the classical limit upto which one can march
% towards $t = 0$ is the Planck time
% $t\approx5.34\times10^{-44}\sec$. All the cosmical parameters shall
% have classical meaning only from $t\approx5.39\times10^{-44}
% \sec$. And the variable $Y$ becomes the classical expression
% \begin{equation*}
% Y = {\biggl(2\sqrt{3} H_{(0)}\biggr)}^{\dfrac{1}{2}}\sqrt{t}
% \end{equation*}

% When $Y$ lies in the range

% \begin{equation*}
% 1.6924079\cdots \le Y < \text { very large number }
% \end{equation*}

% Hence the least value of $ Y = 1.6924079\cdots$. It occurs when
% $t\approx10^{-44}\sec$. (See Appendix II). 

\newpage
\section*{Acknowledgements} 
                    
We are grateful to  Dr. K.S. Vishwanathan (Former Professor and Head
of the Department of Physics, University of Kerala) for personal interest.

\medskip
We are deeply indebted to our friends Sri. Er.M.P. Lokanath, General
Secretary, Internet Society of Kerala and Prof. Eugine Nezarath, Vice
Chairman, Internet Society of Kerala for making use of the Internet
facilities. We also express our deep gratitude to our friends Prof.
R.P. Lalaji, Director, National Institute of Computer Technology,
Kerala and Dr. Rajendra Prasad (Former Professor, T. K. M. Engineering
College, Kollam) for permitting us to make use of computer time.

\renewcommand{\bibname}{References}
%\bibliographystyle{tekunsrt}
%\bibliography{cosmic}

\begin{thebibliography}{1}

\bibitem{one}
Sudhakaran G.
\newblock ``{A} {M}odified form of {E}instein's {F}ield {E}quations''.
\newblock {Abstract of `Global Conference on Mathematical
  Physics' [Centenary Celebration of Niels Bohr and Herman Weyl]}, October 20 - 26, 1987,(Nagpur). Organised by   Einstein Foundation International and Department of Mathematics [Nagpur University], p. 55. 
\newblock Brief account accepted for publication in its proceedings.

\bibitem{two}
Sudhakaran G.
\newblock ``{A} {N}ew {C}oncept of {M}atter''.
\newblock {`Proceedings of National Seminar on Frontiers of Science' (from 15 - 1 - 92 to 17 - 1 - 92) held
  in connection with the 125th Anniversary Celebrations of University College,
  Thiruvananthapuram, Kerala}, pp. 24 -- 26.

\bibitem{three}
Sudhakaran G.
\newblock ``{T}he {P}roblem of {I}nitial {S}ingularity''.
\newblock {`Proceedings of Fourth Kerala Science Congress'(27 - 29 February 1992, Thrissur)} pp. 203 -- 204, 

\bibitem{newfour}
Sudhakaran. G. and Sivaram. C.
\newblock ``Cosmic repulsion in presence of matter''.
\newblock {\em Abstracts of Plenary Lectures and Contributed Papers (GR15
  Conference organsed by IUCAA, Pune), Dec 16 -- 21}, pages 136--137, 1997.

\bibitem{four}
Adler.~R. Bazin.~M. and Schiffer. M.
\newblock {``Introduction to General Relativity''}.
\newblock 2nd Edition, Mc Graw Hill Book Company, 1975,
\newblock p.~337.


\bibitem{five}
Adler.~R. Bazin.~M. and Schiffer. M.
\newblock {``Introduction to General Relativity''}.
\newblock 2nd  Edition, Mc Graw Hill Book Company, 1975,
\newblock p.~409.

\bibitem{six}
Iyer. B.R.
\newblock {``Quantum Field Theory in Curved Spacetime: Cononical
  Quantization'' in Gravitation, Gauge Theories and the Early Universe}.
\newblock Kluwer Academic Publishers, 1989,
\newblock p.~297.

\bibitem{seven}
Adler.~R. Bazin.~M. and Schiffer. M.
\newblock {``Introduction to General Relativity''}.
\newblock Mc Graw Hill Book Company, 2nd Edition, 1975,
\newblock p.~348.

\bibitem{eight}
Donoghue K., Holstein R. and Robinett R.
\newblock {1984 Phys. Rev. D \underline{30}, 2561}.

\bibitem{nine}
Donoghue K.,\textit{et al.}
\newblock {1985 Gen.Rel.Grav.\underline{17}, 207}.

\bibitem{ten}
Donoghue K.,\textit{et al.}
\newblock {1985 Phys. Rev. D \underline{34}, 1208}.

\bibitem{eleven}
Perlmutter S.,\textit{et al.}
\newblock {1997 Ap. J \underline{483}, 565}.

\bibitem{twelve}
Perlmutter S.,\textit{et al.}
\newblock {1998 Nature \underline{391}, 51}.

\bibitem{thirteen}
de Bernardis P.,\textit{et al.}
\newblock {Nature \underline{404}, 955}.

\bibitem{fourteen}
Fiemann J. and Waga I.
\newblock {1998 Phys. Rev. D \underline{57}, 4642}.

\bibitem{fifteen}
Podariu S.,\textit{et al.}
\newblock {2000 Ap. J \underline{532}, 196}.

% \bibitem{sixteen}
% Jou. D. and Pavon. D.
% \newblock Thermodynamics and cosmology.
% \newblock {\em Foundations of Big Bang Cosmology}, page 170. World Scientific,
%   1987.

% \bibitem{seventeen}
% Sivaram, C., Sinha K. P. and Lord E. A.
% \newblock 1974 Nature \underline{249}, 640.

% \bibitem{eighteen}
% Sivaram, C., Sinha K. P. and Sudershan E. C. G.
% \newblock 1976 Found. Phys. \underline{6}, 717 

% \bibitem{nineteen}
% de Sabhata V. and Sivaram, C.
% \newblock 1993 Found. Phys. Lett \underline{6}, 561

% \bibitem{twenty}
% Sivaram, C.
% \newblock 2000 Curr Sci. \underline{79}, 413

% \bibitem{twentyone}
% S. W. Hawking in "Hawking on the Big Bang and Black Holes", (World Scientific, Singapore 1993) p. 242-43.

\end{thebibliography}

\end{document}